# Direct Observation of Phase Oscillation Modes in Josephson-coupled Layered Superconductors

Z. Y. Lai [1, *], X. Yao[2], X. H. Zeng[2], H. Zheng[1], W. Z. Shen[1], X. Y. Chen[1], and H. G. Shen[3]

[1] *Laboratory of Condensed Matter Spectroscopy and Opto-Electronic Physics, Department of Physics, Shanghai Jiao Tong University, 1954 Hua Shan Road, Shanghai 200030, P. R. China*

[2] *Laboratory of High $T_c$ Superconductor, Department of Physics, Shanghai Jiao Tong University, 1954 Hua Shan Road, Shanghai 200030, P. R. China*

[3] *School of Electronics and Electric Engineering, Shanghai Jiao Tong University, 1954 Hua Shan Road, Shanghai 200030, P. R. China*

## ABSTRACT

We have proposed to employ the microwave S parameters method to measure phase oscillation modes (POMs) in Josephson coupled-layered superconductors. For the first time, in $SmBa_2Cu_3O_7$ (SmBCO) and $YBa_2Cu_3O_{7-\delta}$ (YBCO) at 77 K we have directly observed some POMs far below the Josephson plasma frequency. A series of discrete modes have been observed and discussed qualitatively in the POMs. The method may be a powerful and flexible tool in this field.

---

* Corresponding author.
  E-mail: zylai@sjtu.edu.cn

As well known, high $T_c$ superconductor (HTSC) or layered organic superconducting salt may be considered as a stack of superconducting layers coupled by Josephson junction, which is called as intrinsic Josephson junction (IJJ). The novel properties of these materials as compared with a single Josephson junction are associated with this structure and with atomic thickness of the superconducting layers. Theoretically, the physicists have developed some complicated time-dependent equations for the phase difference in different conditions [1-4]. Although general equations describing the phase dynamic are still under discussion, the phase variation in time and space (along c-axis) indicates that there are some phase oscillations in IJJ. Considered the periodic structure along c-axis direction, they should physically consist of a series of oscillation modes, like the simulation by Kleiner *et al.* [5], which hereafter we call as phase oscillation modes (POMs). Obviously, the POMs depend on the coupling between layers, the structure of HTSC, the quasi-particle conductivity, the pining, the charging effect, and some external conditions, such as temperature and external magnetic field. In other words, the POMs certainly contain rich information about the IJJ. Furthermore, it is ineluctable for a lot of POMs to exist in Josephson-coupled layered superconductors. We believe that they would play an important role in the properties of the HTSC. So it is quite important to measure them for understanding the physical properties of the IJJ.

So far, the experiments [6-10] and the theoretical works [1-4, 11, 12] have almost focused on the frequencies and the line-widths of the c-axis Josephson plasma (JP). Some useful information can be extracted [12, 13] from the data about the JP. However, the information is limited because the JP is one kind of the POMs and its frequency, $\omega_0$ (or $\omega_p$ when $\mathbf{B} \neq 0$), is an average result over the temperature and disorder caused by pinning centers [11, 14]. Obviously, much important information on the IJJ was lost in the average. In addition, it is difficult to observe the JP resonances in some superconducting materials because they possess much high JP frequencies beyond present technique.

Except the JP, it is not easy to observe the POMs because the electromagnetic wave in wide frequency range, especially below the JP frequency, is strongly reflected out by the HTSC material in the superconducting state, according to the calculation by Tachiki *et al.*

[15]. However, in order to excite a POM in a HTSC sample, the electromagnetic wave with its frequency has to be applied on the sample. That is why many experiments by virtue of the microwave resonance and absorption have been performed near the JP frequency only [6-10].

To measure the POMs, we suggest a method in terms of microwave S (scattering) parameters, which represent the relationship between the incoming and outgoing waves for a microwave circuit. Up to now, for the first time we have directly observed some modes far below the JP frequency.

Before describing this method, two key points should be noted: (i) although the electromagnetic wave is strongly reflected by a HTSC, the tunneling supercurrent (TSC) of a POM can be excited by the microwave with the same frequency because of resonance, (ii) according to the superconductive theory, the TSC density $j_{n,n+1}$ in a Josephson junction depends on the gauge-invariant phase difference $\varphi_{n,n+1}(\mathbf{r},t)$ in the following way [16, 17]:

$$j_{n,n+1} = j_c \sin\left[\varphi_{n,n+1}(\mathbf{r},t)\right]. \tag{1}$$

This means that in the linear microwave network theory the TSC has to be treated as another current source with the same frequency as the external field in the microwave measurements. The reason is that the relationship between the TSC and the external alternating electric field is nonlinear, that is, it does not follow the usual Ohm's law.

In our method, in order to increase the circuit isolation between the input and output terminals of the measurement set without the sample, we make a U-typed bulk HTSC sample with the c-axis direction shown in Fig. 1(a), where the two microwave strip lines are connected by the sample to constitute a microwave circuit. Then the S parameters of this circuit can be measured with the normalizing resistance $R_0$ (=50 Ω commonly). The circuit can be treated as a Π–like equivalent network shown in Fig. 1(b) when the sample is in the normal state. According to the microwave network theory, there has to be $s_{12}=s_{21}$. However, when the sample is in the superconducting state, the TSC in the IJJ has to be introduced into the network as another current source as mentioned above. Fig. 1(c) shows the equivalent network, where i is the current source, then $s_{12} \neq s_{21}$. Here we define a phase oscillation mode factor (POMF) η, which represents the ratio between $iR_0$ and the voltage of the exciting source E. In terms of the microwave network theory we can obtain:

$$\eta = \frac{2(s_{21} - s_{12})}{(1+2\delta)s_{12} + (1-2\delta)s_{22}}. \qquad (2)$$

Here δ stands for the deviation value from the half coupling impedance of the Π–like equivalent network, as shown in Fig. 1(c), due to the position of the current source. Clearly, if $s_{12} \neq s_{21}$, $\eta \neq 0$, or $i \neq 0$, or there is the TSC in the sample, and vice versa.

In our experiments, the samples are the SmBCO and YBCO single crystals. In manufacturing the SmBCO single crystals, a top-seeded solution growth (TSSG) method was used to grow SmBCO single crystals in air [18]. $Sm_2O_3$ crucibles were used so that the contamination from the crucible could be reduced to a minimum. High-quality $Ba_3Cu_5O_z$ powders were used as raw materials for the solvent. Highly c-axis oriented YBCO thin films were deposited on MgO substrates and used as seeds for growth of SmBCO crystals. The SmBCO and YBCO crystals were cut into two U-typed samples with the size of about 4.55×2.40 and 5.5×2.86 mm$^2$ in the a–b plane, and 3.52 and 4.25 mm in the c-axis direction, respectively. The spaces between the two inner faces of the two arms are about 0.75 and 0.85 mm, and their inner lengths are about 3.03 and 3.0 mm, respectively. The sample was annealed in a tube furnace at 340 $^o$C for 200 h in the oxygen gas flow. $T_c$ of the crystals is about 93 K, tested by superconducting quantum interference devices (SQUID). The substrates of the microwave strip lines are MgO slices with the thickness about 0.50 mm. Entire structure [see Fig. 1(a)] was packed in a copper case. For SmBCO the network analyzer Agilent 8722ES is used to measure the S parameters with the frequency in the range of 5.0 – 25.0 GHz. In order to change the microwave input power conveniently, for the YBCO the network analyzer Anritsu 37369D was used. The results |η| at 300 K and 77 K are respectively shown in Figs. 2(a) and (b) for SmBCO with the microwave input power smaller than -5 dBm and direct current (DC) 0 and 100 mA, and in Figs. 3(a) and (b) for YBCO with 0 mA DC and the microwave power 2, -12, -22, and -27 dBm, respectively. In Figs. 2(a) and 2(b), the curves without DC are shifted up from that with 100mA DC for a clear comparison. As the same reason, the curves corresponding to different microwave input powers are shifted in the Figs. 3(a) and 3(b).

It is obvious that the experimental results of |η| are basically zero at room temperature (300 K), as shown in Figs. 2(a) and 3(a). The lower wavy lines in these figures are due to the noise and system fluctuation. These results are reasonable because the samples are in the normal state at room temperature and there isn't any TSC in the samples. However, in the liquid nitrogen (77 K), there are a series of discrete peaks, as shown in the Figs. 2(b) and 3(b). Considered that the maximum output of Agilent 8722ES is -5dBm, about 0.3 mW, and the input power for YBCO is smaller than 2dBm, about 1.58 mW, our experiments are small-signal measurement. Generally, in this lower microwave power, except the nonlinear property of TSC which is intrinsic and needn't large microwave power to excite, it is impossible that other nonlinear effect, such as harmonic effect, is excited. From the spaces between the peaks in the Figs. 2(b) and 3(b) the change of their amplitudes, and especially almost no changes of the positions and the amplitudes of the peaks in Fig. 3(b) when the microwave input power decreases from 2 dBm to -27 dBm, they are by no means a harmonic effect or an effect of a frequency multiplier. According to the microwave network theory, the condition which makes $s_{12}$ unequal to $s_{21}$ (that is, the POMF η unequal to zero) is that there exists another source in the linear microwave network. In our experimental set in which the direction of the electric field in the microwave is parallel to the c-axis almost, except the TSC with some phase oscillation in the sample which can be excited by the electric field and have been presented as the source as mentioned above, there is no active element which can introduce the source. That is, logically, the peaks in our experiment result are certainly produced by the TSC.

Based on the measurement, we try to analyze our result qualitatively.

Here we use the Josephson's fourth equation [16, 17] to discuss our result with introducing the interaction between layers simply. The equation for gauge-invariant phase difference $\varphi_{n,n+1}(\mathbf{r},t)$ can approximately be written as:

$$\alpha \nabla^2 \varphi - \frac{\varepsilon_r}{c^2} \frac{\partial^2 \varphi}{\partial t^2} = \frac{1}{\lambda_J^2} \sin\varphi. \qquad (3)$$

Here φ is one of $\varphi_{n,n+1}(\mathbf{r},t)$ and α stands for the interaction by other layers. If the φ is small and it can be expressed as $\varphi_s\exp(-i\omega t)$, where $\varphi_s$ is a function of space, from Eq. (3) we obtain

$$\alpha\lambda_J^2\nabla^2\varphi_s=(1-\frac{\omega^2}{\omega_0^2})\varphi_s. \qquad (3a)$$

Here $\nabla$ can be taken as a two-dimension operator because the space between two nearest layers is small so that the variation of $\varphi_s$ along direction z can be neglected. Obviously, according to the Eq. (3a) in the a-b plane the distributions of $\varphi_s$ can be like some standing or damping waves along x and/or y direction as shown in Fig. 1(a) with the factors $\exp(\pm i\beta_\parallel r_\parallel)$ or $\exp(-\beta_\parallel r_\parallel)$ respectively, which we call as standing or damping modes respectively. The $\beta_\parallel$ can be written as:

$$\beta_\parallel^2=\pm\frac{1}{\alpha\lambda_J^2}\left(1-\frac{\omega^2}{\omega_0^2}\right). \qquad (4)$$

Here $\beta_\parallel^2=\beta_x^2+\beta_y^2$. The sign before the term of the right side of this equation is negative for the standing mode and positive for the damping mode, respectively. According to this equation, it is clear that $\omega$ should be larger than $\omega_0$ for standing mode and smaller than $\omega_0$ for damping mode, considered that generally α is positive[11]. That means that it implies below JP there are some phase oscillations with the form of damping wave. So the peaks in Figs. 2(b) and 3(b) indicate that some phase oscillations with damping wave exist in the HTCS because the microwave frequency in our experiment is far below JP.

According to the Eq. (4), we can estimate the order of magnitude of $\beta_\parallel$ in our experiment in terms of the approximate result of the interaction between layers in Ref. [11]. Generally, for our microwave frequency, $\omega_0\gg\omega$. Taken account of almost the same phase between layers due to the damping modes, for one layer the interactions from other layers can approximately be added up numerically. Then $\beta_\parallel\approx s/(\lambda_J\lambda_{ab})$, where s is the distance between nearest two layers. For YBCO or SmBCO s≈1.17 nm，$\lambda_{ab}$≈190 nm, and $\lambda_J$≈16 μm [19], $\beta_\parallel\approx 0.383$ mm$^{-1}$. In terms of the value of $\beta_\parallel$ we can determine the centrobaric position of the distribution of the TSC in the IJJ so that the values of δ can be estimated. For the sample SmBCO δ is about 0.0708 (its size of IJJ along x direction of Fig. 1(a) is about 1.5 mm) and for the sample YBCO about 0.115 (the size about 2.5 mm). Of course, it is difficult to obtain the exact $\beta_\parallel$ for every peak in the Figs. 2(b) and 3(b) now. The reason is: First of all, the general equation describing the phase dynamic is in discussion as mentioned above. Accordingly, how to construct exciting and boundary condition is important.

It is much interested that the phase oscillations corresponding to the peaks in the Figs. 2(b) and 3(b) show a form of some discrete modes. This result is different from that in a single Josephson junction. Obviously, according to the Josephson's fourth equation for single junction the frequency of the phase oscillations, regardless of damping or standing modes, should be continuous. That indicates that the periodic layer structure and its influence on the TSC in the IJJ will play a significant role so that the phase oscillation forms a series of discrete modes, that is, POMs. We think that the electron pairs of the TSC on their passage through the layers suffer the scattering from the periodic layers and the electric field due to some kind of periodic electron distribution on the structure (although the distribution may deviate from equilibrium) creates a periodic potential. By the influence of the potential and the scattering the continuous spectrum of the phase oscillation is broken at some frequencies.

As well known, if the gauge-invariant phase difference is small, it can be expanded as the sum of three parts due to DC, AC, and the external alternating electromagnetic field. So, according to the Eq. (3), when the external field is small, $\varphi$ is proportional to the external field approximately, that is, in our experiment we can suppose it is proportional to 1-R, where R stands for the reflectivity of the sample. Further, for the small amplitude variation of TSC, in term of the Eq. (1), we have:

$$|\eta| \propto i \propto j \propto \Psi_{\alpha n} \propto (1-R). \qquad (5)$$

Figs. 2(c) and 3(c) show the fitting curves of $|\eta|$ for the modes a, b, c, and d in the Fig. 2(b) and a', b' and c' in Fig. 3(b), respectively. In order to obtain the curves, we use the fitting parameters of the curve marked as $T/T_c$=0.8 of Fig. 1 in Ref. [15], which described the $\omega/\omega_p$ dependent of R. The fitting function is a third-order polynomial. The result is satisfactory basically and $\omega_0/2\pi \approx 340$ GHz at 77 K for SmBCO and 500GHz for YBCO.

This value seems reasonable, considered that the SmBCO and YBCO are not a strong asymmetric HTSC.

As observed in Figs. 2(b) and 3(b), in the frequency range of our experiments, in addition to some large peaks there are a series of small peaks between them. The phenomenon can be explained as follow: Among the modes to be excited in our method, different mode has different intensity according to Eq. (5). For example, in our experiments, the microwave transmits in the direction x shown in the Fig. 1(a) so that the main damping modes (MDMs) were along the direction in the a-b plane, that is, the modes should possess the factor $\exp(-\beta_x x)$. Their intensities follow the Eq. (5). The modes a, b, c, and d in the Fig. 2(b) and a', b', and c' in the Fig. 3(b) should belong to the MDM and for sure their intensities follow the equation. However, in our experiment, some modes with the factor $\exp(-\beta_y y)$ can still exist due to impure TEM wave in our microwave strip line, that is, in our structure, the excitation along the direction y in the Fig. 1(a) exists also. Merely the intensity of these modes (we call them as side damping modes (SDMs)) should be weak relative to the MDM nearby. Of course, there are some mixed modes of the MDM and SDM. Their intensity should yet be weak relative to the MDM nearby. The modes f, g, and h in the Fig. 2(b) and d', f', g', and h' in Fig. 3(b) should belong to them. That is why in the estimation of $\omega_0/2\pi$ we selected modes a, b, c, and d, and a', b', and c', respectively. In addition, it is obvious that among all the damping modes many modes may be degenerate. Once external condition is changed, some degenerate modes would be split, or, maybe some new modes be introduced into, or some modes would be moved. This phenomenon can be found in Fig. 2(b) clearly, compared the peaks belonging to 0 mA DC with those to 100 mA DC, for example, at modes c and d.

Besides, our method doesn't contain any resonant cavity. That is, in fact the set to measure POMs can be a wide band microwave device so that many modes can be obtained in an experiment. On the other hand, this method has quite high sensitivity. That can be proved by our experiments, where the mode frequencies are so low relative to the JP that the reflectivity is above about 95%, according to Ref. [15]. Furthermore, the key point of our method is that the TSC must be introduced into a microwave circuit as another current source when S parameters of the microwave devices with the IJJ are measured. In one word, our method may be a powerful and flexible tool in this field.

In conclusion, we have explained the importance to investigate the POMs in the IJJ for studying the properties of the HTSC, and proposed a powerful and flexible method, which can cover wide frequency range and possess high sensitivity, to measure the POMs. For the first time we have directly observed some POMs, in the form of the discrete modes, in the SmBCO and YBCO with their frequencies far below the JP and. Our work may stimulate further studies on the mechanism of phase dynamics in IJJ.

## ACKNOWLEDGMENTS

This work was supported in part by the National Science Foundation of China under the contracts Nos. 10474052, 10125416 and 50272038.

# FIGURE CAPTIONS

Fig. 1 (a) Schematic illustration of the set for measuring POMs in a HTSC. Equivalent microwave network of the set for the HTSC in (b) normal state and (c) superconducting state. i in (c) corresponds to the tunneling super-current in IJJ.

Fig. 2 Amplitude of POMF $|\eta|$ versus $\omega/2\pi$ for SmBCO with the frequency in the range of 5.0-25.0 GHz with input microwave power less than -5 dBm and at (a) T=300 K and (b) 77 K, where the curves corresponding to 0 mA DC are shifted up for clarity. (c) Fitting results of the MDMs a, b, c, and d.

Fig. 3 Amplitude of POMF $|\eta|$ versus $\omega/2\pi$ for YBCO with the frequency in the range of 5.0-26.0 GHz with 0 mA DC and at (a) T=300 K and (b) 77 K, where the curves corresponding to microwave input power 2, -12, and -22 dBm are respectively shifted up for clarity. (c) Fitting results of the MDMs a', b', and c'.

# FIGURES

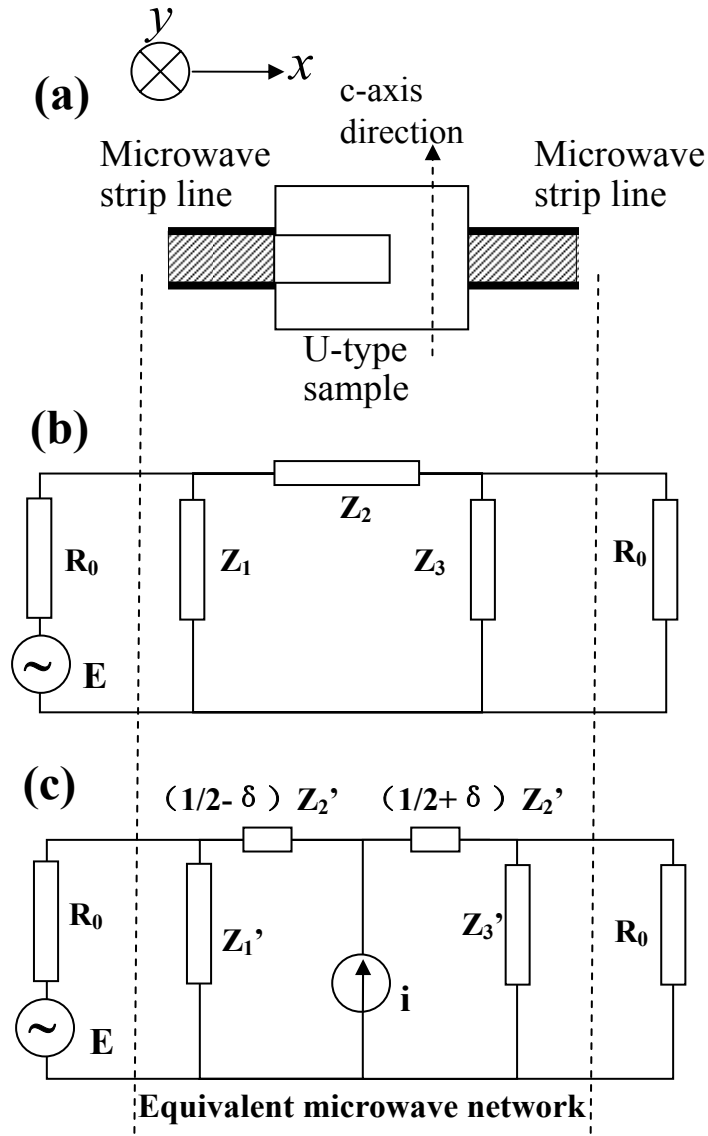

Z. Y. Lai, *et al.*, Fig. 1 of 3

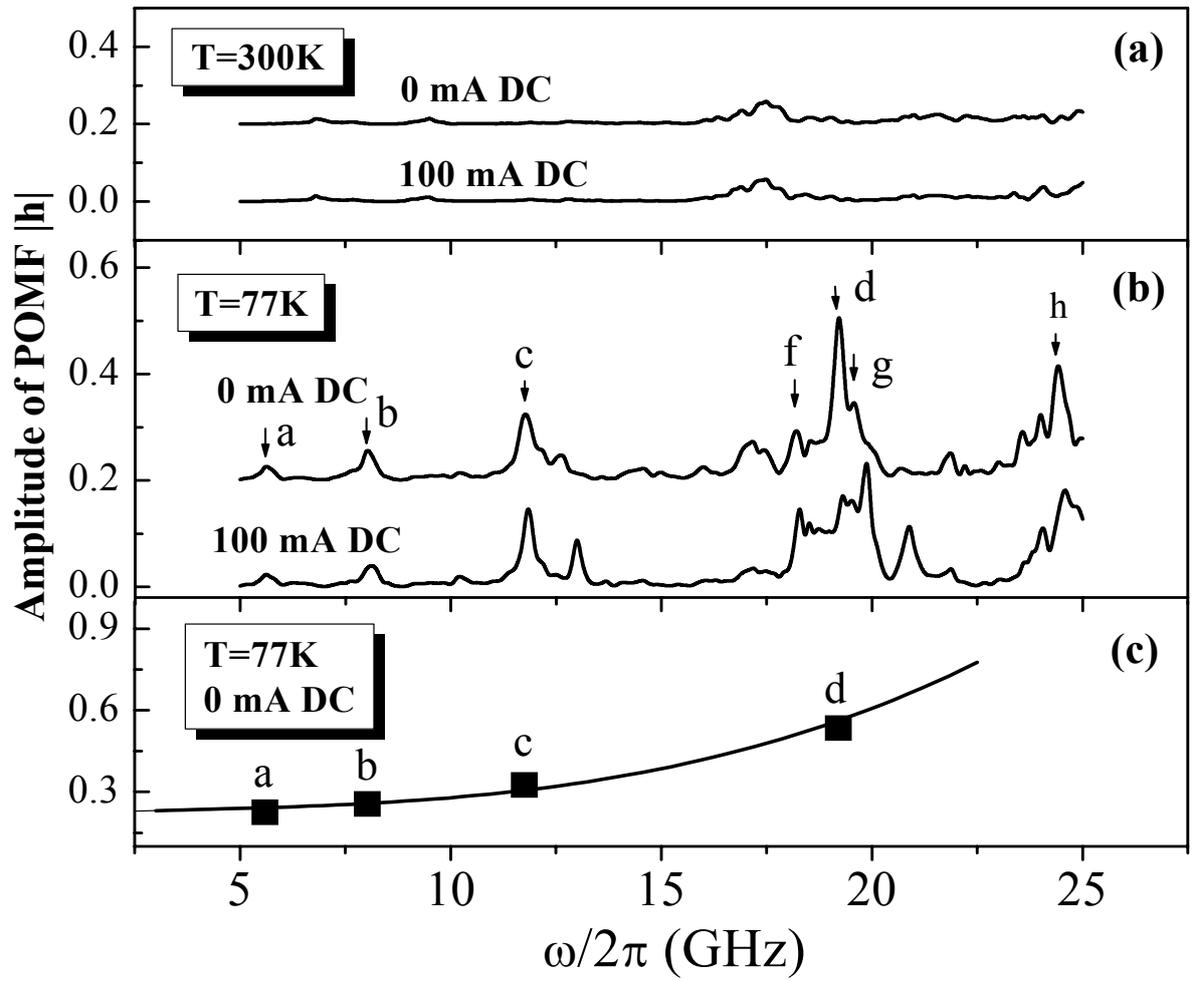

Z. Y. Lai, *et al*., Fig. 2 of 3.

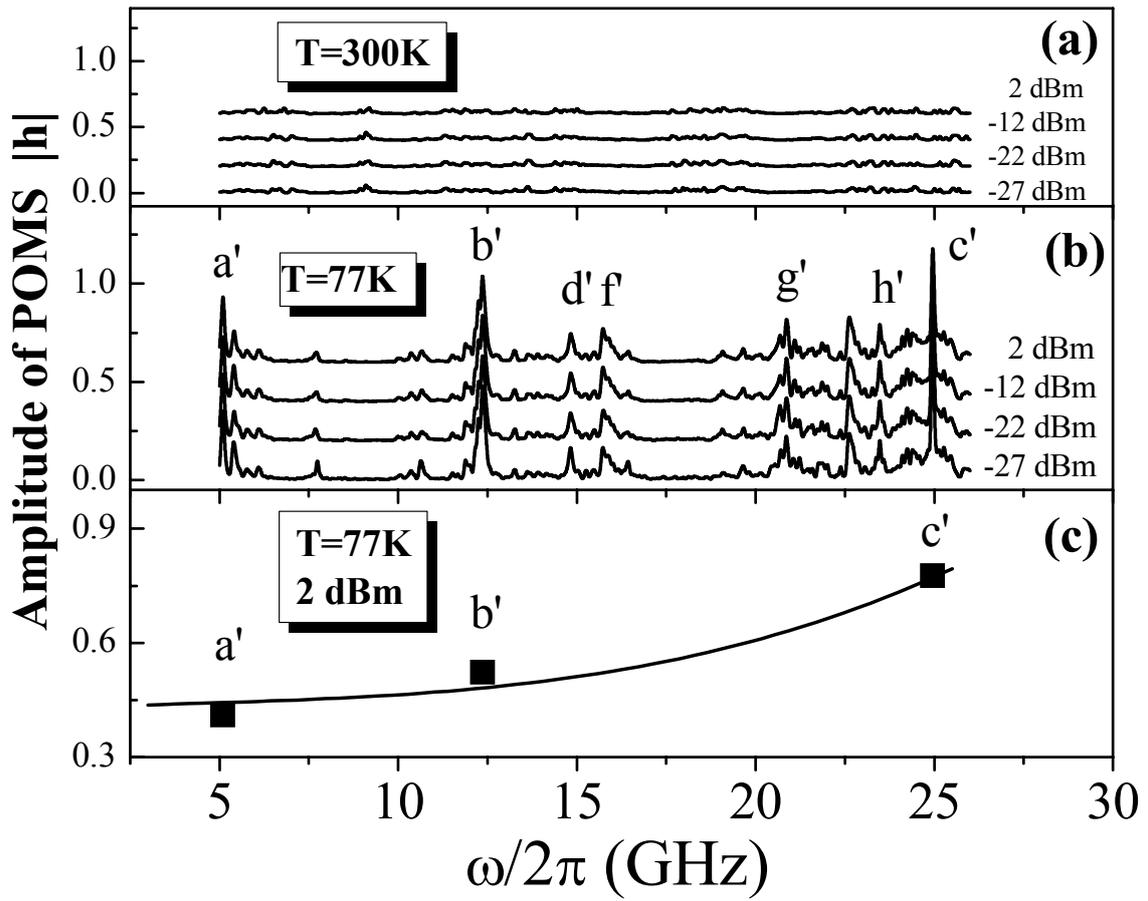